\documentclass[reprint, amsmath, amssymb, prr]{revtex4-2}

\usepackage[T1]{fontenc}
\usepackage[utf8]{inputenc}
\usepackage[english]{babel}

\usepackage{amsmath}
\usepackage{amssymb}
\usepackage{physics}
\usepackage{siunitx}
\DeclareSIUnit\gauss{G}
\usepackage[version=4]{mhchem}

\usepackage{graphicx}
\usepackage{subfig}
\usepackage{float}

\usepackage[capitalise]{cleveref}
\usepackage{xcolor}

\begin{document}

\title{Prospects for single photon sideband cooling\\of optically trapped neutral atoms}

\author{F. Berto}
\email{berto@lens.unifi.it}
\affiliation{Politecnico di Torino, 10129 Torino, Italy}
\affiliation{Istituto Nazionale di Ricerca Metrologica (INRiM), 10135 Torino, Italy}
\affiliation{European Laboratory for Nonlinear Spectroscopy (LENS), 50019 Sesto Fiorentino, Italy}

\author{E. Perego}
\affiliation{Istituto Nazionale di Ricerca Metrologica (INRiM), 10135 Torino, Italy}
\affiliation{European Laboratory for Nonlinear Spectroscopy (LENS), 50019 Sesto Fiorentino, Italy}

\author{L. Duca}
\affiliation{Istituto Nazionale di Ricerca Metrologica (INRiM), 10135 Torino, Italy}
\affiliation{European Laboratory for Nonlinear Spectroscopy (LENS), 50019 Sesto Fiorentino, Italy}

\author{C. Sias}
\affiliation{Istituto Nazionale di Ricerca Metrologica (INRiM), 10135 Torino, Italy}
\affiliation{European Laboratory for Nonlinear Spectroscopy (LENS), 50019 Sesto Fiorentino, Italy}
\affiliation{Istituto Nazionale di Ottica del Consiglio Nazionale delle Ricerche (CNR-INO), 50019 Sesto Fiorentino, Italy}

\date{\today}

\begin{abstract}
We propose a novel cooling scheme for realising single photon sideband cooling on particles trapped in a state-dependent optical potential. We develop a master rate equation from an \emph{ab-initio} model and find that in experimentally feasible conditions it is possible to drastically reduce the average occupation number of the vibrational levels by applying a frequency sweep on the cooling laser that sequentially cools all the motional states. Notably, this cooling scheme works also when a particle experiences a deeper trap in its internal ground state than in its excited state, a condition for which conventional single photon sideband cooling does not work. In our analysis, we consider two cases: a two-level particle confined in an optical tweezer and \ce{Li} atoms confined in an optical lattice, and find conditions for efficient cooling in both cases. The results from the model are confirmed by a full quantum Monte Carlo simulation of the system Hamiltonian. Our findings provide an alternative cooling scheme that can be applied in principle to any particle, e.g. atoms, molecules or ions, confined in a state-dependent optical potential.
\end{abstract}

\maketitle

\section{Introduction}
Optical potentials are widely used for confining and controlling particles like atoms~\cite{Bloch2005, Muldoon2012}, molecules~\cite{Ospelkaus2006, Anderegg2019}, and, more recently, ions~\cite{Schneider2010, Perego2020}.
Individual particles can be reliably isolated and arranged in optical potentials, resulting in ensembles that represent a hardware for quantum computation and quantum simulation. Celebrated examples are optical lattices~\cite{Bloch2012} and arrays of optical tweezers~\cite{Pagano2019}, in which the spatial ordering of the particles can be precisely controlled.

In all these applications, particles must be efficiently cooled in order to increase their lifetime, observe quantum effects arising at low energies, and reduce the effects of decoherence. 
Laser cooling of particles confined in optical potentials is particularly challenging, since the trap depth is typically comparable to the Doppler temperature associated with the laser cooling transition of most atomic elements.
Therefore, in order to reach lower temperatures, sub-Doppler cooling techniques are employed, e.g. sideband cooling~\cite{Neuhauser1978}. 

Sideband cooling is based on the selective laser excitation of motional quantum states of a trapped particle~\cite{Diedrich1989}. 
This is possible when the energy separation between the motional levels is larger than the recoil energy associated with the cooling transition, a condition that is fulfilled in the so-called Lamb-Dicke regime. 
In this regime, the particle excitation spectrum exhibits sidebands, each of which associated with the transfer of an integer number of motional quanta from the light field to the particle. 
Sideband cooling is achieved when one or more lasers addressing a cooling transition are used to excite a sideband associated to a lower motional state. 
After the excitation, dissipation is realised by the spontaneous emission of a photon to the internal ground state. 
Remarkably, sideband cooling can reduce the particles' motional state well below the Doppler limit: \emph{e.g.}, in the case of ions in radiofrequency traps, a ground state occupation larger than \( 99.9\% \) has been achieved~\cite{Roos1999}.

However, with respect to its realisation on ions in radiofrequency traps, sideband cooling of particles confined in an optical trap has an additional complication: the optical potential is proportional to the complex polarizability \(\alpha\) associated to the particle's state~\cite{Grimm2000}, and therefore the potential experienced by the particle in the two levels of the cooling transition is in general different. 
When this is the case, the resonant frequency for the cooling transition is dependent on the particle motional state, thus a particle can be resonant to different laser frequencies depending on its kinetic energy. A possible solution to this problem is to use optical potentials at ``magic'' wavelengths, so that the polarizabilities of the ground state \(\alpha_g\) and of the excited state \(\alpha_e\) are equal. However, these ``magic'' wavelengths can be, depending on the particle, at inconvenient values for their practical realisation in an experiment.
An alternative strategy is to perform sideband cooling by using two-photons Raman transitions, so that the levels used for sideband cooling can be two sublevels (e.g. hyperfine or Zeeman) of the electronic ground state manifold~\cite{Groebner2017}. 
However, this scheme requires the use of several wavelengths, whereas a single photon sideband cooling remains a desirable solution for efficiently implementing laser cooling in optical traps.

Theoretical models of sideband cooling in and out of the Lamb-Dicke regime have been rigorously developed for both atoms and ions~\cite{Javanainen1981, Stenholm1986, Morigi1999}. 
A more general model including the case in which the optical trapping potential depends on the internal state of the trapped particle has also been developed~\cite{Taieb1994}.
In this study, the authors point out that cooling mainly originates from the difference in the potential energy accumulated during the motion by a particle in the internal ground or excited state, i.e. a ``Sisyphus-like'' effect that is independent from the momentum transfer between the particle and the photons. 
The authors conclude that cooling can occur only in case \(\alpha_g \le \alpha_e\), in which the Sisyphus effect is dissipative. 
However, this limitation puts a strong bound for the experimental implementation of the optical potential. 
A recent experimental study~\cite{Cooper2018} observed that, in the case of \( \alpha_g > \alpha_e \), cooling is still possible using a laser blue detuned with respect to the free atom transition frequency. 
Cooling is obtained by realising a ``Sisyphus cap'', i.e. a condition in which a laser is resonant to the cooling transition when the particle has potential energy \( E_\text{cap} \). 
As a result, a particle with energy \( E \le E_\text{cap} \) is cooled, and a particle with \( E > E_\text{cap} \) is heated up and eventually lost from the trap. However, this process causes losses that affect the efficiency of the cooling scheme. 

In this work, we propose a novel scheme for implementing single-photon sideband cooling in optical traps that surpasses current limitations by making it possible to cool particles in any optical potential, and in particular for \(\alpha_g > \alpha_e\).
This scheme is based on sweeping the frequency of the cooling laser that ensures the cooling of all motional levels, leading to a final energy close to the potential ground state. 
We provide a full quantum treatment of the cooling process, and show its performance within the two-level system (TLS) approximation by considering two systems having the masses and transition linewidths of \ce{Li} and \ce{Yb}. 
This allows us to simulate the cooling process on particles having considerably different characteristics. Additionally, we present a possible experimental implementation of the cooling process considering \ce{Li} atoms trapped in an optical lattice, in which we take into account the contributions from all the levels of the \ce{Li} hyperfine manifold. 
We verify the results of our analytic model by using a quantum Monte Carlo numerical simulation and find excellent agreement.

This paper is organised as follows. 
In Sec.~\ref{sec:model}, we develop the mathematical model for sideband cooling of atoms trapped in a non-harmonic potential with a state dependent depth. 
In Sec.~\ref{sec:tweezers}, we present a numerical investigation of the model within the two-level system approximation by considering \ce{Li} and \ce{Yb} atoms loaded in a single dipole trap with gaussian intensity profile (optical tweezer). 
In Sec.~\ref{sec:optical_lattice}, we investigate sideband cooling of \ce{Li} atoms loaded in an optical lattice by addressing the D1 transition and by considering the whole multi-level structure of both electronic ground and excited states.
\section{The model}\label{sec:model}
Let us consider a two-level particle with a transition linewidth \( \gamma \) confined in a 1D optical potential \( V \) that is most generally dependent on the atomic polarizability of the particle's internal state. 
The particle is coupled to a near-resonance cooling laser propagating along the z-axis, with electric field \( \vb{E}_\text{c} = \vb{\epsilon}\, E_0^\text{(c)} \cos(k_\text{c} z - \omega_\text{c} t) \), where \( \vb{\epsilon} \) is the light polarisation, \(E_0^\text{(c)}\) the field amplitude, \( k_\text{c} \) the laser wavevector, and \( \omega_\text{c} \) the laser frequency. 
In the direct coupling formalism, the system Hamiltonian is written as

\begin{equation}
	H = H_0 + V - \vb{d}\cdot\vb{E}_\text{c}\,,
	\label{eq:TLS_hamiltonian}
\end{equation}
where \( \vb{d} \) is the dipole operator, and \( H_0 \) the Hamiltonian of the free particle. For the moment, the only assumption that we make on the optical potential \(V\) is that it is a continuous and differentiable function, \emph{e.g.} an optical lattice or an optical tweezer. We also assume that the Hamiltonian in \cref{eq:TLS_hamiltonian} is separable, so we can reduce to the one-dimensional problem (along the z-axis) without loss of generality.
In order to fully quantize \cref{eq:TLS_hamiltonian} in one dimension, we introduce the basis \( \{ \ket{i, n_i} \}= \ket{i} \otimes \ket{n_i} \text{ where the index } i = g, e\)  indicates the ground and the excited internal state of the particle, respectively, and \( n = 1, \dots, n_{\text{i,max}} \) indicates the particle motional level, where \( n_{\text{i,max}} \) is the last bound level of the potential. The states \( \{\ket{n_i}\} \) are eigenstates of \( V \).

Working in this basis the fully quantized form of \cref{eq:TLS_hamiltonian} is given by

\begin{equation}
\begin{aligned}
	H &= \sum_{n_g} \qty(E_{g, n_g} + \frac{\hbar\omega_c}{2})\! \ketbra{g, n_g}{g, n_g} +\\
	&\quad + \sum_{m_e} \qty(E_{e, m_e} - \frac{\hbar\omega_c}{2})\! \ketbra{e, m_e}{e, m_e} +\\
	&\qquad + \frac{\hbar\Omega}{2}\!\! \sum_{n_g, m_e}\!\! \qty[M_{m_e, n_g}\!\ketbra{g, n_g}{e, m_e} + \text{H.c.}]\,,
	\label{eq:TLS_hamiltonian_quantized}
\end{aligned}
\end{equation}
where \( \Omega = \matrixel{e}{\vb{d}\cdot \vb{E}_\text{c}}{g} \) is the Rabi frequency, \( M_{m_e, n_g} = \matrixel{n_g}{e^{ik_\text{c} \hat{z}}}{m_e}\), and \(E_{g, n_g}\) \( (E_{e,m_e})\) is the energy of the particle in the ground (excited) internal state and the \(n_g\) \((m_e)\) level of the optical potential.
We note that in writing \cref{eq:TLS_hamiltonian_quantized}, we performed the rotating wave approximation and we are considering a transformed wavefunction rotating in the same frame of the cooling laser.

In order to take into account the spontaneous emission, which itself may cause changes in the TLS motional state~\cite{Javanainen1981}, we introduce the Lindblad operators

\begin{equation}
	L_{m_e, n_g} = \sqrt{\gamma_{m_e, n_g}} \ketbra{g, n_g}{e, m_e}\,,
	\label{eq:lindblad_ops}
\end{equation}
with rates \( \gamma_{m_e, n_g} \) given by

\[
	\gamma_{m_e, n_g} = \frac{\gamma}{2} \int_{-1}^{+1} \dd{u} N(u) \abs{\matrixel{n_g}{e^{iku\hat{z}}}{m_e}}^2.
\]
The integration over \( u \) averages over all possible direction of the emitted photon wavevector \(k\) weighted by the dipole emission angular distribution \( N(u) = 3(1+u^2)/8 \). For \( u = \pm 1 \) the photon is emitted along the z-axis, in this case the momentum transfer is maximal, on the other hand \( u = 0 \) indicates a photon emitted orthogonal to the z-axis and no change in the motional state along the z-axis occurs.\\
The system dynamics is described by the Lindblad equation

\[
	\dot{\rho} = -i[H, \rho] + \sum_{j} L_{j}\, \rho\, L_{j}^{\dag} - \frac{1}{2} \qty{L_{j}^\dag\, L_{j}, \rho}\,,
\]

where \( \rho \) is the density operator and \( j = (n_g, m_e) \) runs over all bound states indices.

It is possible to obtain a master rate equation describing the occupation of different bound levels by performing an adiabatic elimination of the excited states. 
The condition for performing the adiabatic elimination is that the coupling between ground and excited states is weak, \emph{i.e.} the natural decay dynamics must be much faster than any other dynamics of the system~\cite{Brion2007}, resulting in a small occupation of the excited state. 
In the sideband cooling model developed by Stenholm \emph{et al.}~\cite{Stenholm1986} -- for which the potential experienced by the particle is independent from the particle's internal state -- this requirement is satisfied for the sidebands in the Lamb-Dicke regime, i.e. if \( \eta \ll 1 \), where \(\eta = a_{0} k_\text{c}\) is the Lamb-Dicke parameter and \(a_{0}\) is the spatial extension of the lowest bound state of the potential.
However, as pointed out by Ta\"{i}eb \emph{et al.}~\cite{Taieb1994}, the Lamb-Dicke condition is not sufficient for performing the adiabatic elimination if the trap potential depends on the TLS internal state, as in this case the additional requirement of a low intensity of the cooling laser, i.e. \( \Omega / \gamma \ll 1 \), must be verified. 
Under this assumption, we can adiabatically eliminate the excited states following the procedure presented by Reiter and Sørensen~\cite{Reiter2012}, and decouple the diagonal and the off-diagonal terms of the density matrix. The resulting rate equation for the trap populations \( \Pi_{n_g} = \expval{\rho}{g, n_g} \) is

\begin{widetext}
\begin{equation}
	\dot{\Pi}_{n_g} = - \sum_{m_e, h_g} \abs{\gamma_{m_e, h_g}\!}\, \abs{M_{n_g, m_e}\!}^2\ \Gamma\qty(\delta_{m_e, n_g})\, \Pi_{n_g} + \sum_{h_g} \qty(\sum_{m_e} \abs{\gamma_{m_e, n_g}\!}\, \abs{M_{h_g, m_e}\!}^2\ \Gamma\qty(\delta_{m_e, h_g})) \Pi_{h_g}\,,
\label{eq:TLS_master_rate_eq}
\end{equation}
\end{widetext}
where we have introduced the scattering rate function \( \Gamma(\delta) = {(\Omega/2)}^2 / (\delta^2 + \gamma^2 / 4) \) and the detuning \( \delta_{m_e, n_g} = (E_{e, m_e} - E_{g, n_g}) / \hbar - \omega_\text{c} \).\\
The first, negative term on the right-hand side of \cref{eq:TLS_master_rate_eq} represents the loss of population from the state \( \ket{g, n_g} \) due to transitions to the excited states \( \ket{e, m_e} \) followed by spontaneous decay to \( \ket{g, h_g} \), while the second term represents the increase of the \(\ket{g,n_g}\) population due to an absorption-spontaneous emission cycle starting from the \(\ket{g,h_g}\) level.

The matrix elements \( M_{m_i, n_j} \) can be approximated, in the Lamb-Dicke regime \( \eta \ll 1 \), as

\[
\begin{aligned}
	\matrixel{n_j}{e^{ik\hat{z}}}{m_i} &\approx I_{m_j, n_i} +\\
	&\quad + i\,\eta\,\qty[(m_j +1)\, I_{m_j+1, n_i} + m_j\, I_{m_j-1, n_i}]\,,
\end{aligned}
\]

where \( I_{m_j, n_i} = \braket{n_i}{m_j} \) are the overlap integrals of the spatial wavefunctions of the particle in the excited \(\ket{m_j}\) and ground \(\ket{n_i}\) internal levels. In our analysis, we compute these integrals numerically. \cref{eq:TLS_master_rate_eq} can be written as a matrix first-order differential equation

\[
	\dot{\mathbf{\Pi}} = \mathbf{A} \cdot \mathbf{\Pi}
\]
where \( \mathbf{A} \) is the coefficient matrix and \( \mathbf{\Pi} \) is the column vector, the elements of which are the populations \( \Pi_{n_g} \). 
If \( \mathbf{A} \) is time-independent, the general solution is a linear combination of eigenvectors of \( \mathbf{A} \). 
In general, since the coefficients appearing in \cref{eq:TLS_master_rate_eq} are not time-independent, one may take a dense partition of time and approximate \( \mathbf{A} \) as constant over each (small) time interval. Finally, the solution is given by diagonalizing \( \mathbf{A} \) over each time interval, thus approximating the actual solution with a step function.

\begin{figure}
	\centering
	\includegraphics[width=\columnwidth]{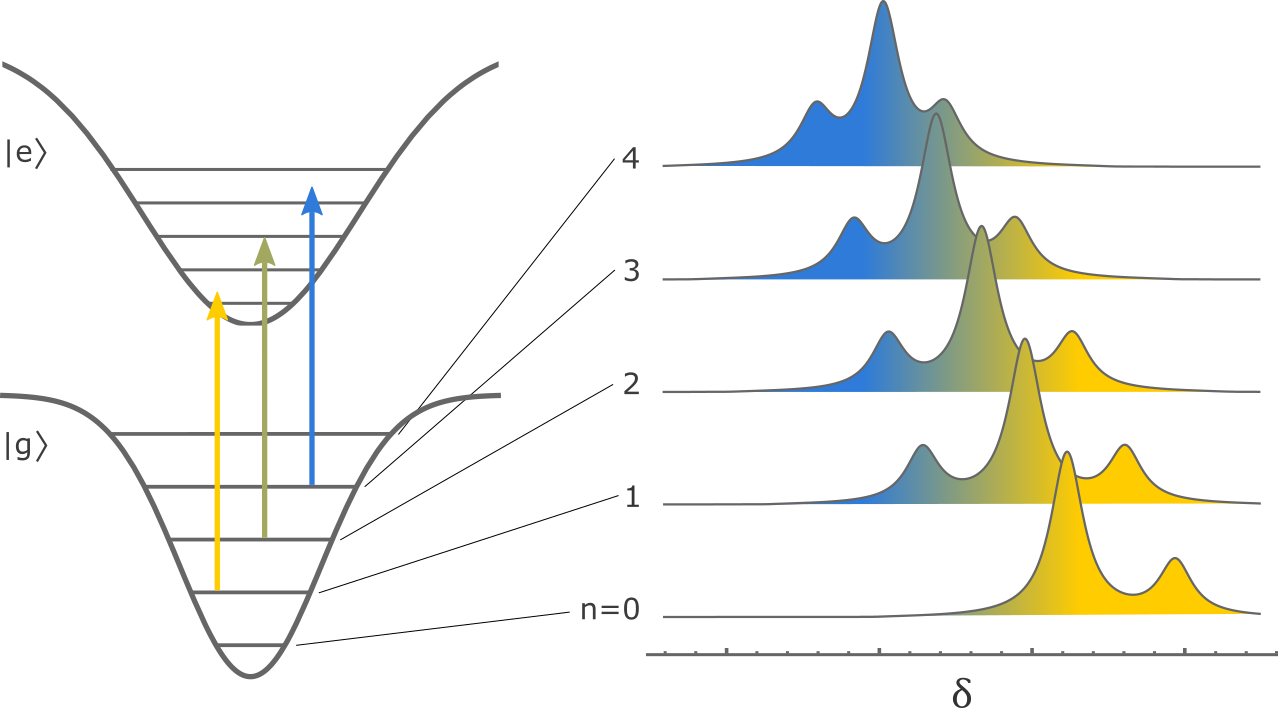}
	\caption{Sketch of the spectrum of a particle confined in a state-dependent optical potential. Different atomic polarizabilities result in different trap depths for the ground and the excited states (left drawing, the case \( \alpha_g > \alpha_e \) is shown). As a result, the excitation spectrum depends on the motional state of the particle (right plot, only carrier and first sidebands transitions are shown). In this scenario a laser resonant with the carrier transition \( n=2 \to n'=2 \) (represented by the arrows) will excite mainly the particle's blue sidebands for \( n \ge 4 \) (blue regions of the spectrum), and red sidebands for \( n < 3 \) (yellow regions of the spectrum).}\label{fig:trap_spectrum}
\end{figure}

To generalise the model to a multilevel system, in particular to the hyperfine structure of an atom, we introduce the quantum numbers \( F \) and \( m_F \) representing hyperfine levels in the ground state, and \( F' \) and \( m_F' \) for the excited hyperfine states. Then one can perform the following substitutions in \cref{eq:TLS_master_rate_eq}: \( n_g \to (F, m_F, n_g) \), \( m_e \to (F', m_F', m_e) \) and \( h_g \to (F'', m_F'', h_g) \), where \( F'' \) and \( m_F'' \) are quantum numbers indicating a ground hyperfine state different than \( F \) and \( m_F \). Note that in general \( \gamma \) and \( \omega \) will also depend on the particular \( (F, m_F),\ (F', m_F') \) combination considered~\cite{SteckBook}.

As we already pointed out, the motional energy spectra of a particle in the ground and in the excited states are in general different. Thus, the detuning (with respect to the unperturbed transition) of carrier and sidebands transitions will generally depend on the initial motional state~\footnote{For sideband cooling of atoms in an harmonic potential the frequency position of carrier and sidebands is independent of initial (or final) motional states, as the spacing between these is constant and the trap potential is independent of its internal state.} and will spread out over a large frequency range, as shown in \cref{fig:trap_spectrum}. 
This suggests that, as predicted by Ref.~\cite{Taieb1994}, in case \(\alpha_{g}>\alpha_{e}\) cooling with a single laser frequency is inefficient, since a laser resonant to the red sideband of the \( n \)-th level will scatter mainly photons from the blue sideband of the \( l \)-th levels with \( l > n \), causing heating.
However, in the next two sections we will demonstrate that efficient cooling can be restored when the model includes a linear sweep of the cooling laser frequency, \( \omega_c(t) \). 
The frequency of the cooling laser, initially resonant with the free space atomic transition, is swept towards larger detunings in order to sequentially cool the lower energy levels. In this way, the laser scans all the red sidebands starting from the high energy motional levels down to approximately the ground state, thus favouring the cooling of the largest possible number of atoms.

\section{Sideband cooling in optical tweezers}
\label{sec:tweezers}

We consider the case of two different TLSs, having the transition frequencies and the masses of \ce{Li} and \ce{Yb}.
We choose these atomic species as they have considerably different atomic linewidths and nuclear masses, thus making it possible to study the cooling process under different but still experimentally feasible conditions. 
With respect to \ce{Yb}, the two-level approximation is particularly well suited for the bosonic isotopes for which there is no hyperfine splitting. 
For what concerns the \ce{Yb} case, we consider an optical tweezer implemented with a single laser beam at \SI{532}{\nano\metre} with a waist of \SI{700}{\nano\metre} and \SI{25.8}{\milli\watt} power, resulting in a trap depth of \( V_{0,g} = \SI{2.1}{\milli\kelvin} \) for the \ce{^1S_0} ground state and \( V_{0,e} = \SI{1.7}{\milli\kelvin} \) for the \ce{^3P_1} state. 
Cooling is performed on the \( \ce{^1S_0}\to\ce{^3P_1} \) intercombination line having a linewidth of \( 2\pi\cross\SI{182}{\kilo\hertz} \). 
In the case of \ce{Li}, we consider an optical tweezer realised with a laser of wavelength \SI{1064}{\nano\metre}, waist \SI{700}{\nano\metre} and \SI{47.5}{\milli\watt} power, resulting in trap depths \( V_{0,g} = \SI{3.7}{\milli\kelvin} \) and \( V_{0,e} = \SI{2.6}{\milli\kelvin} \) for the \( \ce{^2S_{\flatfrac{1}{2}}} \) and \(\ce{^2P_{\flatfrac{1}{2}}} \), respectively. 
We consider the \( \ce{^2S_{\flatfrac{1}{2}}}\to\ce{^2P_{\flatfrac{1}{2}}} \) D1 line as the cooling transition, having a linewidth of \( 2\pi\cross\SI{5.8}{\mega\hertz} \). 
The depths of these traps were chosen such that for both atoms the ground state Lamb-Dicke parameter \( \eta_g \) is equal to \num{0.2}. The traps' waist and power are comparable with the parameters of experiments with optical tweezers reported in literature~\cite{Yu2018, Saskin2019}. Furthermore we note that for both atomic species the equivalent harmonic trapping frequency \( \omega_T \) is smaller than the transition linewidth \(\gamma\).

\begin{figure}
	\centering
	\includegraphics[width=\columnwidth]{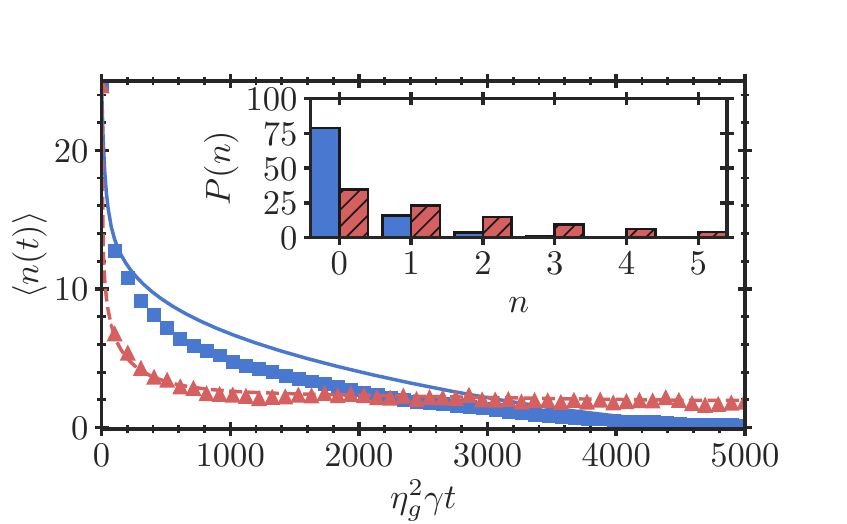}
	\caption{Average occupation number as a function of time during sideband cooling of  \ce{Yb} (blue solid line) and \ce{Li} (red dashed line). The data are obtained by numerical integration of \cref{eq:TLS_master_rate_eq}. The blue squares and the red triangles indicate the results from a quantum Monte Carlo simulation for \ce{Yb} and \ce{Li}, respectively. The time axis has been scaled by the sideband scattering rate \( \eta_g^2 \gamma \)~\cite{Stenholm1986}. Inset shows the occupation probability of the lowest bound levels after cooling of \ce{Yb} (blue bars) and \ce{Li} (red hatched bars).\label{fig:tweezer_results}}
\end{figure}
We simulate the cooling process by numerically integrating \cref{eq:TLS_master_rate_eq} while the cooling laser frequency is slowly changed linearly in time and the laser intensity is set to \( 0.1\, I_\text{sat} \), where \(I_\text{sat}\) is the saturation intensity of the cooling transition. The calculation was truncated to the first \(60\) bound levels of the optical potentials, as for higher energy levels the calculation of the particle wavefunction becomes computationally demanding. In order to have a non-negligible population in the energy levels that were computed, we considered for the initial state a thermal distribution with an average occupation number of \(60\), corresponding to a temperature of \SI{0.4}{\milli\kelvin} for \ce{Yb} and \SI{2.6}{\milli\kelvin} for \ce{Li}. The distribution was then truncated to the first \(60\) levels and normalised to 1.


In a first simulation, we calculate the mean particle's motional number \( \expval{n(t)} \) after performing a sweep of the cooling laser. The results of the simulation are reported in \cref{fig:tweezer_results}. 
We observe that the average occupation number is reduced over time for both \ce{Li} and \ce{Yb} atoms, until the particles reach a steady mean energy level. The ground state occupation after the laser sweep is \SI{80.0}{\%} for \ce{Yb} and \SI{34.6}{\%} for \ce{Li}, the energy reduction is \SI{98.2}{\%} and \SI{92.0}{\%} for the two species, respectively.  The sweep durations are \SI{110}{\milli\second} for \ce{Yb} and  \SI{3.4}{\milli\second} for \ce{Li}, corresponding in both cases to \(\eta_{g}^2\gamma t = 5000\). 
The sweep start frequency corresponds to the red sideband of the least bound state considered, while the sweep stop frequency is optimized in order to avoid resonant heating that is triggered once the \(\ket{g,0}\to\ket{e,0}\) carrier frequency is excited. In spite of the fact that the product between the red sideband linewidth \( \eta_g^2 \gamma\) and the laser sweep time are kept equal in the simulation, the dynamic behaviour of the occupation number of \ce{Yb} and \ce{Li} is qualitatively different. In particular, \ce{Li} atoms reach the minimum attainable energy after approx. \(1/3\) of the sweep time. We attribute this discrepancy to the different values of the ratio \( \omega_T/\gamma \), which are \num{0.57} and \num{0.126} for \ce{Yb} and \ce{Li}, respectively.
The occupation probability distribution after the cooling process is a thermal distribution (see inset \cref{fig:tweezer_results}), and the associated temperatures are \SI{4.3}{\micro\kelvin} and \SI{114}{\micro\kelvin} for \ce{Yb} and \ce{Li}, respectively.



In order to validate our results, we performed a full quantum Monte Carlo simulation using \cref{eq:TLS_hamiltonian_quantized} and \cref{eq:lindblad_ops}. The simulation is implemented using the QuTiP framework~\cite{Johansson2013}. The laser sweep is accounted for by changing \( \omega_c \to \omega_c(t) \) in the system Hamiltonian. 
The initial quantum state of each particle is randomly selected with weights given by the initial occupation distribution of the trap levels. 
We do not simulate the center-of-mass motion along the weakly confining axis of the tweezer. 
With the previously indicated trap and sweep parameters, the resulting dynamical behaviour of the occupation number averaged over 250 trajectories is shown in \cref{fig:tweezer_results}. We observe a reduction of the particles' motional energy, with a dynamics that qualitatively agrees with the numerical solution of the master rate equation. In particular, the final mean occupation number of the ground state potential is in excellent agreement with the solution of the master rate equation.  
We note that the Monte Carlo simulation is performed directly on the Hamiltonian~(\ref{eq:TLS_hamiltonian_quantized}) and, therefore, it does not require the adiabatic elimination of excited states.

\begin{figure}
	\centering
	\includegraphics[width=\columnwidth]{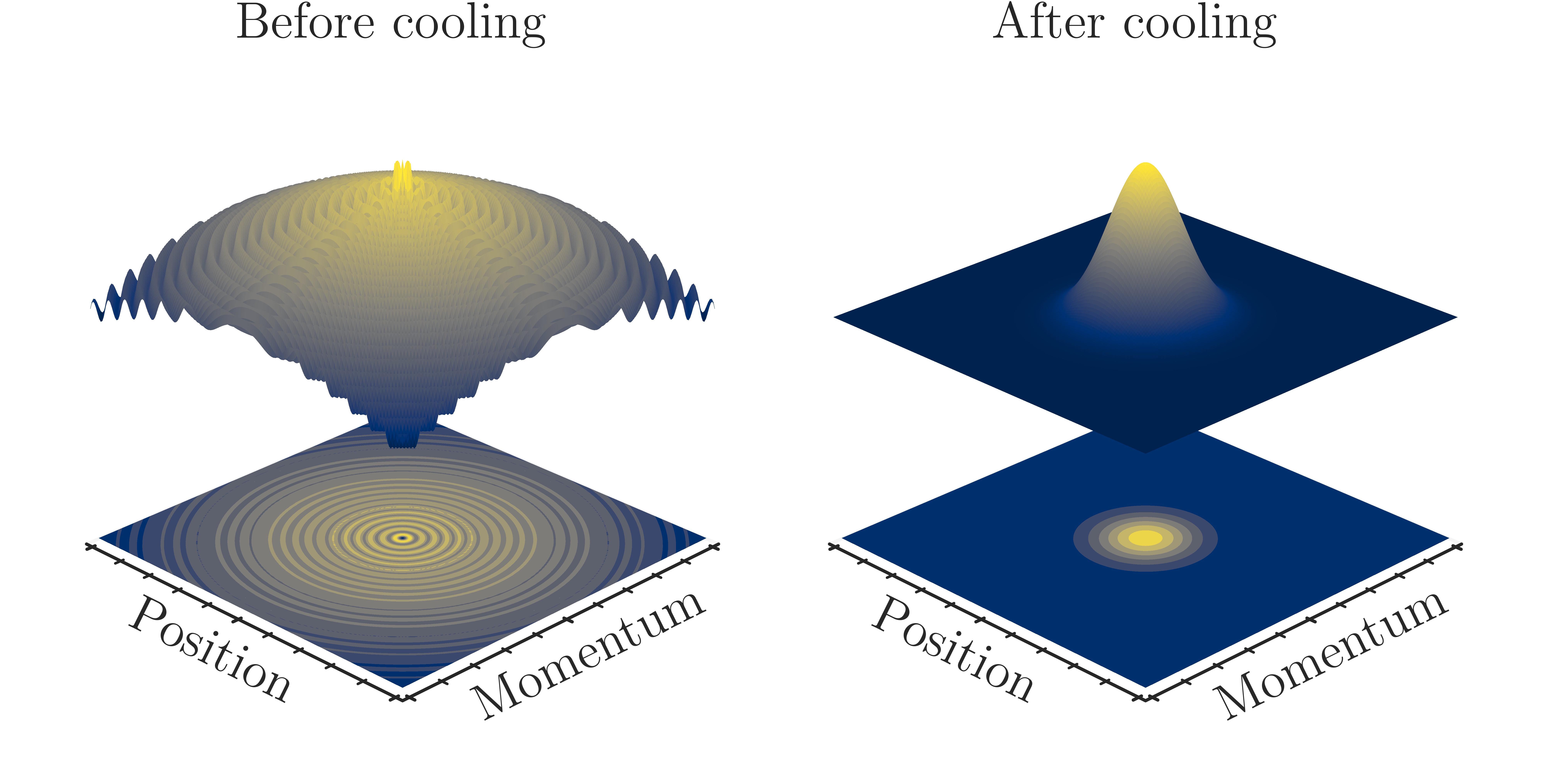}
	\caption{Wigner quasiprobability distribution before and after cooling of \ce{Yb}. Vertical axes have been scaled for better visualisation.}\label{fig:wiger_func}
\end{figure}
\cref{fig:wiger_func} shows the Wigner quasiprobability distribution before and after the cooling process in the case of \ce{Yb} atoms.
We find that initially the atom is broadly spread in phase space (\cref{fig:wiger_func}, left plot) indicating a relatively hot particle. 
At the end of the laser sweep, the distribution in the phase space is compressed near the origin (\cref{fig:wiger_func}, right plot), showing a reduction of particle momentum and occupied positions, thus demonstrating the cooling of the particle's motion. 
This final distribution has a gaussian shape and it is similar to the occupation probability of a Fock state with \( n=0 \). 
Both distributions are centered on the origin as expected for a particle in a symmetric potential. 
Using the computed Wigner function, we observe that the variance of position and momentum of the particle reduces linearly during cooling (not shown), which might indicate that the cooling process could be optimised by using a nonlinear sweep, \emph{e.g.} with by using optimal quantum control~\cite{Caneva2011}.

\begin{figure}
	\centering
	\includegraphics[width=\columnwidth]{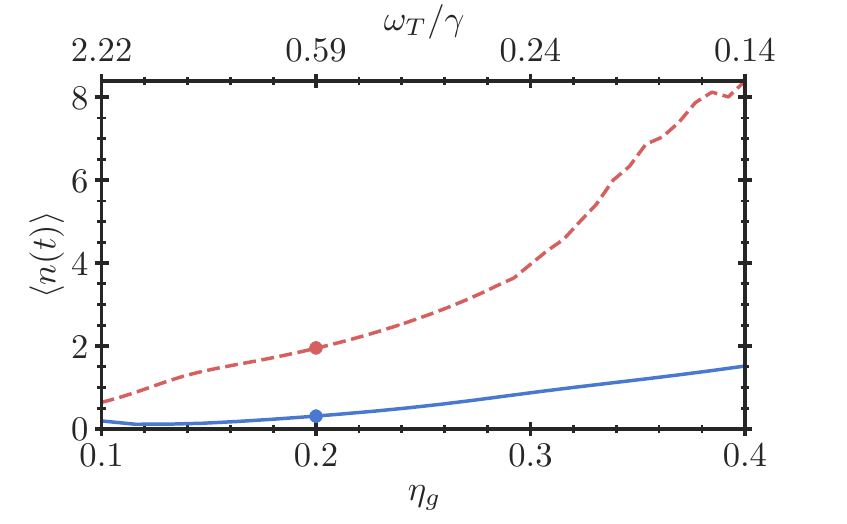}
	\caption{Average occupation number after the frequency sweep of the cooling laser as a function of the ground state Lamb-Dicke parameter for \ce{Yb} (blue solid line) and \ce{Li} (red dashed line). The x-axis above the plot reports the values of the ratio \( \omega_T / \gamma \) for \ce{Yb}, the corresponding values for \ce{Li} can be obtained by rescaling the axis by the ratio of the atomic linewidths. Solid dots indicate the parameters used in the simulations of \cref{fig:tweezer_results}. Different values of \( \eta_g \) are obtained by changing the trapping laser power.}\label{fig:tweezer_VARETA}
\end{figure}
The efficiency of the cooling process is strongly dependent on the choice of the experimental parameters. Therefore, we study the resulting occupation number as we vary different parameters: the trap depth, the difference in polarizability and the cooling laser frequency sweep time.
\cref{fig:tweezer_VARETA} shows the mean occupation number after cooling for different values of the Lamb-Dicke parameter \( \eta_g \), which is changed by varying the trap depth. As a result, also the ratio \( \omega_T / \gamma \) is varied (top x-axis).
We find that the cooling efficiency is considerably reduced when \( \eta_g \) is increased, as expected since the motional sidebands are less resolved when the condition for the Lamb-Dicke regime is relaxed and the ratio \( \omega_T / \gamma \) is increased. The reduction in cooling efficiency is more pronounced in the case of \ce{Li}. We attribute this to the fact that the \( \omega_T / \gamma \) ratio is far smaller for \ce{Li} with respect to \ce{Yb} due to the different atomic linewidths.

\begin{figure}
	\centering
	\includegraphics[width=\columnwidth]{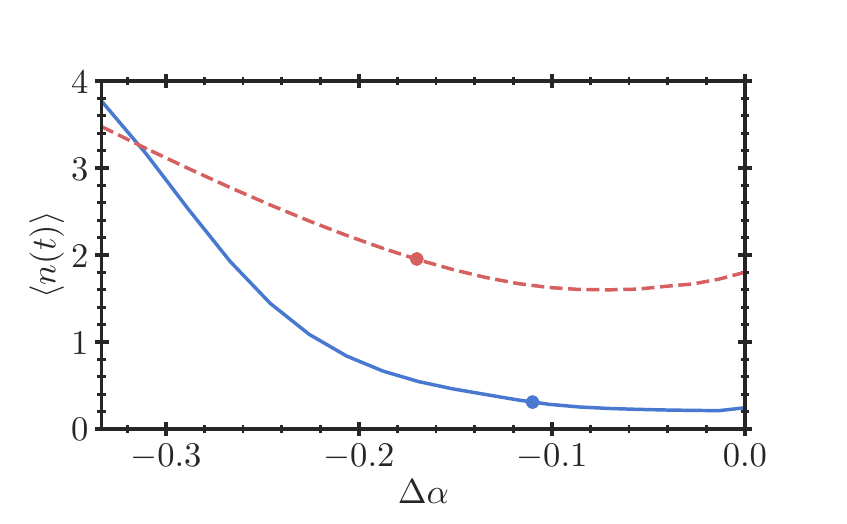}
	\caption{Average occupation number after cooling as a function of the difference in polarizability \( \Delta\alpha \) for \ce{Yb} (blue solid line) and \ce{Li} (red dashed line). \( \Delta\alpha=0 \) implies a magic trapping condition, while negative values are associated with a trap for which a particle in its internal ground state is more bound than a particle in the excited state. Solid dots indicate the parameters used in the simulations of \cref{fig:tweezer_results}. Different values of \( \Delta\alpha \) are obtained by changing the atomic polarizability for the excited state only.}\label{fig:tweezer_VARPOL}
\end{figure}
The difference in polarizability determines the difference between the trap depths experienced by an atom in the ground or in the excited internal state. Therefore, this quantity determines the spread of the sidebands' transition frequencies. 
As the energy difference between the carrier and the sidebands increases, the coupling strength between levels with different motional state decreases, resulting in a lower efficiency of the cooling process. 
By artificially changing the atomic polarizability of the excited state, we explore how the occupation number after the laser sweep depends on the difference in polarizability \(\Delta\alpha=(\alpha_e - \alpha_g)/(\alpha_e + \alpha_g)\). The results of this simulation are shown in \cref{fig:tweezer_VARPOL}. As predicted, the further away the system is from the magical trapping condition, the lower is the cooling efficiency.  
Cooling of \ce{Yb} is more efficient than \ce{Li} except for the region where \( \Delta\alpha < -0.3 \), in which the two curves cross each other. We attribute this effect to the different linewidths \(\gamma\) of the cooling transitions. In fact, while the frequency of a harmonic trap scales with \(m^{-1/2}\), in this specific case the ratio between the cooling linewidths \(\gamma_{\ce{Li}}/\gamma_{\ce{Yb}}\) is larger than the scaling factor of the harmonic trap frequencies \((m_{\ce{Li}}/m_{\ce{Yb}})^{1/2}\). As a consequence, with respect to the cooling transition linewidth, the transitions \(\ket{n_g}\to\ket{m_e}\) for different values of \(\ket{n_g}\) are resolved in \ce{Yb} atoms at a value of \(\abs{\Delta\alpha}\) lower than in \ce{Li} atoms, resulting in a worsening of the cooling efficiency. In support of this interpretation, we performed a separate simulation in which we arbitrarily set the linewidth of the \ce{Li} cooling transition equal to the \ce{Yb} one. We found that an increase of \(\abs{\Delta\alpha}\) results in an increase of the mean occupation number \(\expval{n(t)}\) that is larger for a lighter particle, as expected from the scaling of the harmonic trap frequencies. In the analysis of \cref{fig:tweezer_VARPOL}, we also note that for \ce{Li} cooling is more efficient when \(\Delta\alpha\) is slightly smaller than one. We attribute this effect to the relatively large value of the \(\gamma/\omega_{tr}\) ratio in \ce{Li}, which results in a non-negligible excitation of the carrier transitions during the cooling process. This effect is reduced if \(\Delta\alpha\lesssim0\), since this causes a small mismatch between the transition frequencies \(\ket{n_g}\to\ket{m_e}\) for different values of \(n_g\).



\begin{figure}
	\centering
	\includegraphics[width=\columnwidth]{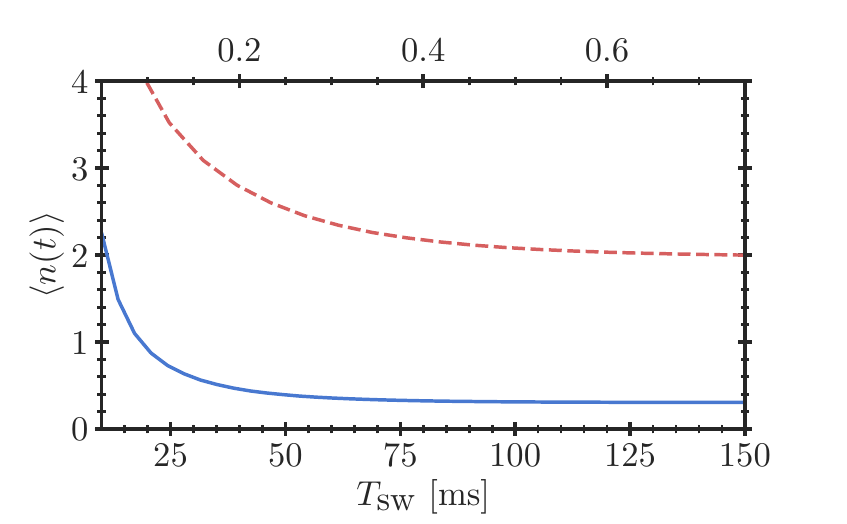}
	\caption{Average occupation number after cooling as a function of the sweep duration \(T_{\text{sw}}\) for \ce{Yb} (blue solid line) and \ce{Li} (red dashed line). The lower (upper) axis indicates the sweep duration for \ce{Yb} (\ce{Li}).}\label{fig:tweezer_VARTIME}
\end{figure}
Finally, \cref{fig:tweezer_VARTIME} shows the cooling efficiency as a function of the frequency sweep duration when the start and stop frequencies are kept constant.
A threshold-like effect is observed if the laser sweep time is gradually reduced. We attribute this to the fact that, if the laser spends insufficient time at resonance with a given sideband transition, there will be a reduced population transfer between bound levels, which limits the reduction of vibrational quanta.

\section{Sideband cooling of Lithium in an optical lattice}\label{sec:optical_lattice}
We now consider the case of multi-level fermionic \ce{Li} atoms trapped in a one-dimensional optical lattice. 
\ce{Li} atoms in an intense optical lattice are a physical system for which Raman sideband cooling has been extensively studied, e.g.\ in quantum gas microscope experiments~\cite{Blatt2015, Parsons2015}. 
In our simulation, we consider the potential created at the center of a Fabry-Perot optical cavity with finesse \( \mathcal{F} = 17000 \), a resonant mode of waist \( \SI{116}{\micro\metre} \) and a coupled laser of wavelength \SI{1064}{\nano\metre} and power \SI{300}{\milli\watt}. The \( \ce{^2S_{\frac{1}{2}}} \) and \( \ce{^2P_{\frac{1}{2}}} \) levels experience a trapping potential depth of \( \num{8900}E_\text{rec} \) (\SI{17.4}{\milli\kelvin}) and \( \num{6300}E_\text{rec} \) (\SI{12.3}{\milli\kelvin}), respectively. We assume an initial temperature of \SI{40}{\micro\kelvin}, as this is the temperature that can be reached with \ce{Li} in a grey molasses~\cite{Burchianti2014}. The trapping laser polarisation is parallel to the quantization axis (\(\pi\)-polarized)~\footnote{In the following we will always consider a right-handed reference frame with the \( z \)-axis aligned to the optical lattice wavevector, and the quantization axis is orthogonal to it.}.
Under these assumptions, the lattice potential can be written as 

\begin{equation}
	V_{i}(z) = V_{0}^{i} \cos^2(k_\text{tr} z)\,,
	\label{eq:lattice_potential}
\end{equation}
where \( k_{\text{tr}}=2\pi/\lambda_{\text{tr}} \) is the wavevector of the optical lattice at wavelength \( \lambda_{\text{tr}} \) and \( V_0^{i} \) is the trap depth experienced by an atom in the ground (\(i=g\)) or the excited (\(i=e\)) state. 
For a particle moving in a potential as in \cref{eq:lattice_potential}, the energy levels and the corresponding wavefunctions are the solutions of a Mathieu equation~\cite{Connor1984}. We compute the initial trap bound levels populations by assuming that the atoms are instantaneously transferred from an initial free-particle state, assumed to be a gaussian wavepacket of size \( \approx \SI{1}{\milli\metre} \), to the trap bound states. The resulting distribution is peaked near the lowest energy levels with a tail extending toward higher energy bound states. The initial average occupation number is \(\approx 22.7\). We initialise all hyperfine levels with equal population, \emph{i.e.} the atomic cloud is not in a polarised state.

\begin{figure}
	\centering
	\includegraphics[width=\columnwidth]{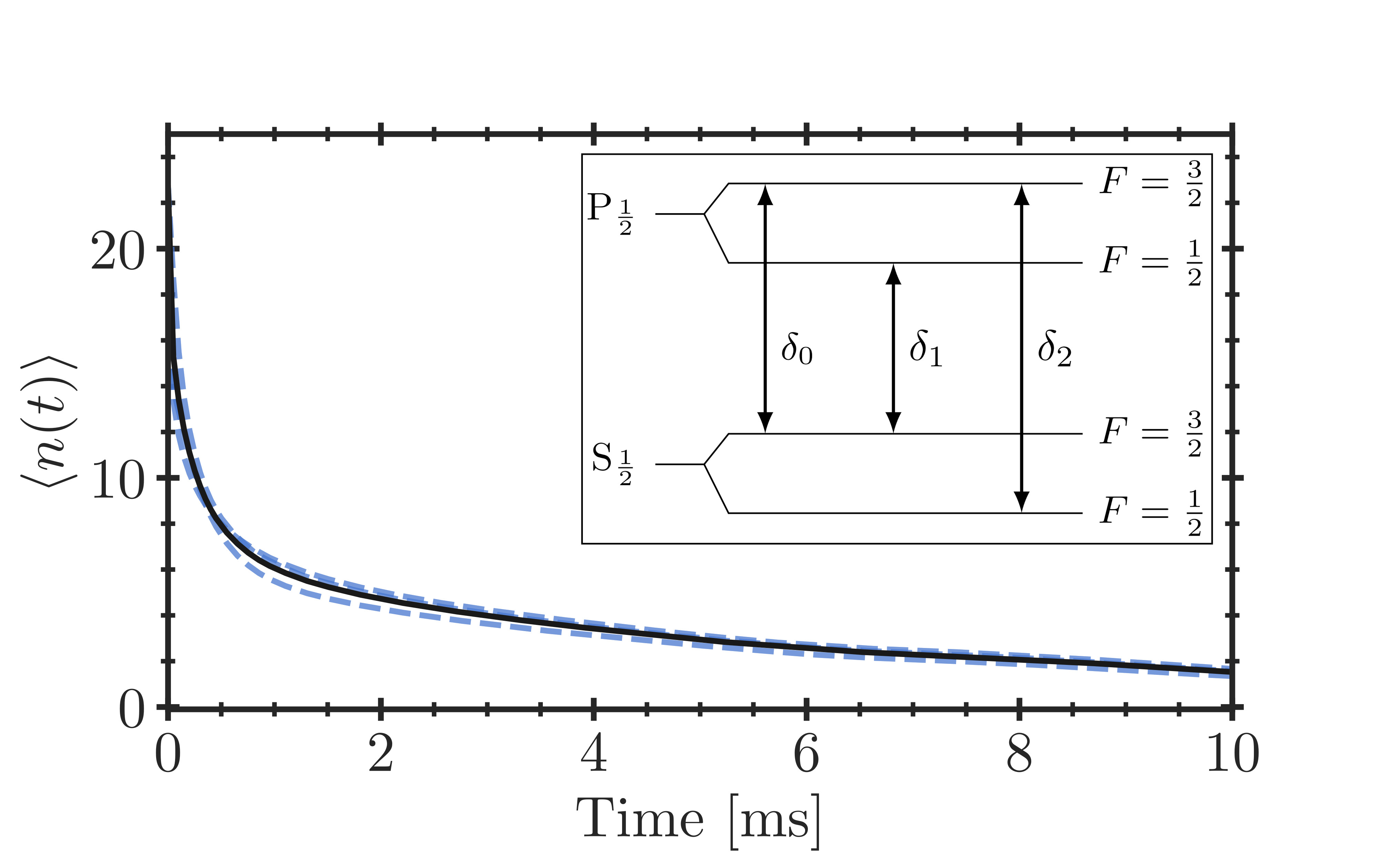}
	\caption{Average occupation number of Li atoms in the \(F=3/2\) (red dashed lined) and \(F=1/2\) (blue dotted line) states as a function of time. The black solid line shows the average of the hyperfine levels. Inset: internal levels of the electronic ground state of \ce{^6Li}. The arrows indicate the transitions that were used in the simulation.}\label{fig:hfs_results}
\end{figure}
We consider the case of a low magnetic field, for which the interactions can be neglected, and for which most hyperfine states are nearly degenerate. We find that the system can be efficiently cooled by using at most three laser frequencies (the addressed transitions are shown in the inset of \cref{fig:hfs_results}) swept together. 
The light detunings with respect to the fine structure \( D_1 \) transition of Lithium are set to: \( \delta_0 \) from \SI{-67.35}{\mega\hertz} to \SI{-12.65}{\mega\hertz}, \( \delta_1 \) from \SI{-93.45}{\mega\hertz} to \SI{-38.75}{\mega\hertz} and \( \delta_2 \) from \SI{160.8}{\mega\hertz} to \SI{215.5}{\mega\hertz}. 
We numerically solve the rate equation and compute the average occupation number \( \expval{\Pi_{F, m_F, n_g}(t)}{F, m_F, n_g} \) for each hyperfine state. 
Results are shown in \cref{fig:hfs_results}. We note that the curves corresponding to hyperfine states with the same \( F \) number are nearly identical. The final occupation number (averaged over different hyperfine levels) is \num{1.53}, with a corresponding ground state occupation probability of \SI{28.3}{\%}. Given these sweep detunings, at higher magnetic fields we observe that cooling is no longer efficient for some of the hyperfine states, in particular when the magnetic field is increased over \SI{20}{\gauss}. 

\begin{figure}
	\centering
	\includegraphics[width=\columnwidth]{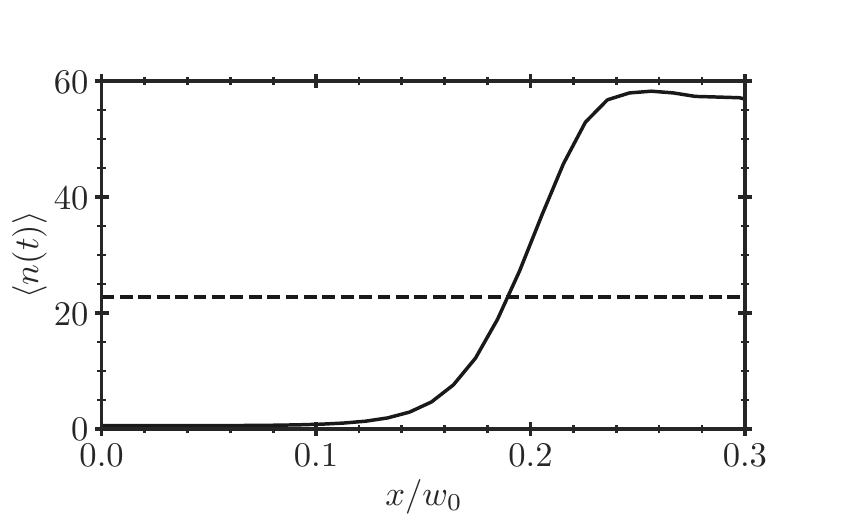}
	\caption{Final average occupation number as a function of the atoms' position along the trap radial axis \(x\). Dashed line indicates the initial average occupation number. The x-axis is in units of the trap laser waist \( w_0 \). Due to the reduction of the trap depth out of the central region, only atoms near the center of the potential are efficiently cooled.}\label{fig:tls_offcentre}
\end{figure}
The bound levels' energies (and thus the sideband transitions) are also dependent on the radial position of the atom in the optical trapping potential. Generally, in an atomic cloud confined in an optical lattice the atoms do not lay exactly on the trap axis where the potential depth is maximal. 
In order to estimate the effects of the radial distribution of a particle within a single lattice site, we simulate the cooling process of an atom displaced at a distance \(x\) from the trap axis.
\cref{fig:tls_offcentre} shows the results of this simulation, in which we use the same sweep parameters as before. 
Due to the different bound state configuration, the laser sweep can only cool efficiently the atoms located near the trap axis while atoms located at a distance greater than \( \geq \SI{10}{\%} \) of the trap waist will experience heating. Assuming that the radial density of the atomic cloud follows a Boltzmann distribution, we calculate for an initial temperature of \( T=\SI{40}{\micro\kelvin} \) a width of the radial density distribution of approximately \SI{10}{\micro\metre}, i.e. \(8.6\%\) of the optical lattice laser waist. 
Therefore, under these realistic conditions cooling is not affected by the finite size of the atomic cloud.

\section{Conclusions}
We have presented a novel cooling scheme based on single-photon sideband cooling of trapped particles in a non-harmonic optical potential with state-dependent trap depth. 
The scheme makes use of a linear frequency sweep of the sideband cooling laser, and we find that efficient cooling is also possible in a trap for which \( \alpha_g > \alpha_e \), \emph{i.e.} for which a particle in its ground state is more bound than in its excited state. Notably, in this condition single-photon sideband cooling with a laser at a fixed frequency was demonstrated to be impossible. 
We simulated the dynamics of a particle by introducing a master rate equation model, and by performing a quantum Monte Carlo simulation of the system Hamiltonian. We have first considered the case of two different TLSs (with the masses and linewidths of \ce{Yb} and \ce{Li}) confined in an optical tweezer. 
Simulations show that it is possible to considerably reduce the average occupation number of the motional state to a value close to the ground state. 
We have also presented a possible experimental application of sideband cooling of fermionic \ce{Li} trapped in a cavity-enhanced standing wave optical lattice. Numerical results show that with three lasers swept in frequency in a range of few tens of \si{\mega\hertz} it is possible to reduce the average vibrational number to a fraction of its initial state in a few \si{\milli\second}.

This general model suggests that similar results can be obtained using different trapping schemes and different atomic systems. 
Moreover, the efficiency of the cooling scheme might be further increased by using non-linear sweeps, \emph{e.g.} by making use of optimal quantum control routines.
Our work is relevant for all experimental realisations of optical trapping of particles like atoms, molecules and ions, and in particular in experiments of quantum simulation and computation, for which cooling of the atomic sample is essential.

\section{Acknowledgement}
We thank M. Zaccanti, G. Roati, and F. Scazza for helpful discussions, and M. Inguscio for continuous support. We thank G. Bertaina for support in the use of computational resources at INRiM. The authors wish to acknowledge that the simulation has been performed with the open-source scientific python framework~\cite{Virtanen2020}. 
This work was financially supported by the ERC Starting Grant PlusOne (Grant Agreement No.
639242), and the FARE-MIUR grant UltraCrystals (Grant No. R165JHRWR3).

\bibliography{main}

\end{document}